\documentclass[a4paper,11pt]{article}
\usepackage{amsmath}
\usepackage{amssymb}

\begin{document}




\thispagestyle{empty}
\begin{titlepage}

\vspace*{-1cm}
\hfill \parbox{3.5cm}{BUTP-97/35 \\ hep-ph/9711431} \\
\vspace*{1.0cm}

\begin{center}
  {\large {\bf \hspace*{-0.2cm}NOTE on the MASS SQUARE of the ETA$^\prime$
  MESON}
      \footnote{Work
      supported in part by the Schweizerischer Nationalfonds.}  }
  \vspace*{3.0cm} \\

{\bf
     P. Minkowski} \\
     Institute for Theoretical Physics \\
     University of Bern \\
     CH - 3012 Bern , Switzerland
   \vspace*{0.8cm} \\  

22. November 1997

\vspace*{4.0cm}

\begin{abstract}
\noindent
We propose to use as a mathematical tool to study the mass square of 
the eta$^{\ \prime}$ meson an infinitely heavy supplementary quark flavor Q.
This is understood in the framework of QCD with one light quark flavor 
($\rightarrow \ n_{\ fl}$ light flavors) and the gauge group 
$SU3_{\ c}$ ($\rightarrow \ SUN_{\ c}$).
The full system thus consists of 
$n^{\ (+)}_{\ fl} \ = n_{\ fl} \ + \ 1 $ flavors, including the infinitely
heavy one. The purpose of this note is to show how the heavy flavor
can be made to represent the anomalies of the subtheory describing the remnant 
physical degrees of freedom. In the large $N_{\ c}$ limit the mass square 
of eta$^{\ \prime}$ tends to a finite limit. The {\em essential} deviations 
from (semi) perturbative derivations are related to simple properties 
of the 'heavy' flavor.
\end{abstract}
\end{center}

\end{titlepage}







\pagestyle{plain}




\noindent
{\bf 1. Central anomalies, single local operator insertions}
\vspace*{0.1cm} 

\noindent
We first turn to the trace anomaly \cite{trace} to demonstrate the
method of using a limiting 'heavy' auxiliary flavor of quark and antiquark.
\vspace*{0.1cm} 

\noindent
All quantities which include the auxiliary flavor shall be marked with the superfix
$^{\ (+)}$. The operator identity for the trace of the energy momentum
tensor with 

\begin{equation}
  \label{eq:1}
n^{\ (+)}_{\ fl} \ = n_{\ fl} \ + \ 1
    \hspace*{0.3cm} ; \hspace*{0.3cm} n_{\ fl} \ \geq \ 1
\end{equation}

\noindent
then takes the form

\begin{equation}
\begin{array}{l}
  \label{eq:2}
\left ( \ \vartheta^{\ (+)} \ \right )^{\ \mu}_{\ \mu} \ = \ - \ b_{\ 1}^{\ (+)}
\ {\cal{B}}^{\ 2}
\hspace*{0.2cm}
      \begin{array}{c} 
        1
        \vspace*{0.2cm} \\
        \hline \vspace*{-0.2cm} \\
        8 \ \pi^{\ 2}
      \end{array}
\hspace*{0.2cm}
+ \ m_{\ q} \ \overline{q} \ q \ +
\ m_{\ Q} \ \overline{Q} \ Q
        \vspace*{0.5cm} \\
{\cal{B}}^{\ 2} \ =
\ {\cal{B}}^{\ 2} \ ( \ N_{\ c} \ ) \ =
     \ \left[
     \ \frac{1}{4} 
     \  F_{\mu \nu}^{a}
        F^{a\, \mu \nu}
     \ \right]
    \hspace*{0.3cm} ; \hspace*{0.3cm}
b_{\ 1}^{\ (+)} \ = \ \frac{11}{3} \ N_{\ c} \ - \ \frac{2}{3} 
\ n^{\ (+)}_{\ fl}
\end{array}
\end{equation}

\noindent
All operators in eq. (\ref{eq:2}) are local and renormalization group invariant 
with vanishing anomalous dimension. In particular the field strengths
forming the bilinear composite operator ${\cal{B}}^{\ 2}$ include the
coupling constant as multiplicative factor, relative to their
perturbative variants. For details of the precise definitions we refer to 
\cite{trace} and \cite{MLPM}. Furthermore ${\cal{B}}^{\ 2}$ does not 
depend on the number of quark flavors.
$ - \ b_{\ 1}^{\ (+)}$ stands for the first coefficient of the (relative)
Callan-Symanzik function $\beta^{\ (+)} \ ( \ g \ ) \ 16 \ \pi^{\ 2} \ / \ g^{\ 3}$.
\vspace*{0.1cm} 

\noindent
Appropriate sums over flavor and color are understood in the definition of
the light flavor mass term $m_{\ q} \ \overline{q} \ q$. The flavor basis
including the 'heavy' Q is chosen such, that the (scheme dependent)
quark masses are all nonnegative. The 'heavy' flavor limit corresponds to
$m_{\ Q} \ \rightarrow \ \infty$.
\vspace*{0.1cm} 

\noindent
In parallel with the operator $\left ( \ \vartheta^{\ (+)} \ \right )^{\ \mu}_{\ \mu}$
we consider its counterpart $\left ( \ \vartheta \ \right )^{\ \mu}_{\ \mu}$,
excluding from consideration the 'heavy' quark flavor.
The corresponding trace identity then reads :

\begin{equation}
\begin{array}{l}
  \label{eq:3}
\left ( \ \vartheta \ \right )^{\ \mu}_{\ \mu} \ = \ - \ b_{\ 1}
\ {\cal{B}}^{\ 2}
\hspace*{0.2cm}
      \begin{array}{c} 
        1
        \vspace*{0.2cm} \\
        \hline \vspace*{-0.2cm} \\
        8 \ \pi^{\ 2}
      \end{array}
\hspace*{0.2cm}
+ \ m_{\ q} \ \overline{q} \ q
    \hspace*{0.3cm} ; \hspace*{0.3cm}
b_{\ 1} \ = \ \frac{11}{3} \ N_{\ c} \ - \ \frac{2}{3} 
\ n_{\ fl}
        \vspace*{0.3cm} \\
\left ( \ \Delta \ \vartheta \ \right )^{\ \mu}_{\ \mu} \ =
\ \left ( \ \vartheta^{\ (+)} \ \right )^{\ \mu}_{\ \mu}
\ - \ \left ( \ \vartheta \ \right )^{\ \mu}_{\ \mu} \ =
\hspace*{0.2cm}
      \begin{array}{c} 
       {\cal{B}}^{\ 2}
        \vspace*{0.2cm} \\
        \hline \vspace*{-0.2cm} \\
        12 \ \pi^{\ 2}
      \end{array}
\hspace*{0.2cm}
+ \ m_{\ Q} \ \overline{Q} \ Q
\end{array}
\end{equation}

\noindent
The key point involves the vanishing of the operator 
$\left ( \ \Delta \ \vartheta \ \right )^{\ \mu}_{\ \mu}$ defined in 
eq. (\ref{eq:3}), when inserted into Green functions pertinent to the
subtheory with only $n_{\ fl}$ flavors of quark and appropriately
restricted momenta (or distances) in the 'heavy' flavor limit
$m_{\ Q} \ \rightarrow \ \infty$.
\vspace*{0.1cm} 

\noindent
We illustrate this considering the two gluon matrix element to lowest
nontrivial order

\begin{equation}
\begin{array}{l}
  \label{eq:4}
  \left \langle \ \varepsilon_{\ 2} \ , \ a_{\ 2} \ , \ k_{\ 2} \ \right |
\ \left ( \ \Delta \ \vartheta \ \right )^{\ \mu}_{\ \mu}
\ \left | \ \varepsilon_{\ 1} \ , \ a_{\ 1} \ , \  \ k_{\ 1} \ \right \rangle \ =
        \vspace*{0.3cm} \\
\delta_{\ a_{2} \ a_{1}} 
\ \varepsilon_{\ 2}^{\ * \ \sigma_{\ 2}} \ \varepsilon_{\ 1}^{\ \sigma_{\ 1}}
\ \left \lbrack \ g_{\ \sigma_{2} \ \sigma_{1}} \ ( \ k_{\ 2} \ k_{\ 1} \ )
\ - \ ( \ k_{ 1} \ )_{\ \sigma_{\ 2}} \ ( \ k_{ 2} \ )_{\ \sigma_{\ 1}}
\ \right \rbrack 
\ \Pi \ + \ ...
        \vspace*{0.3cm} \\
\Pi \ =
\ \Pi \ ( \ k_{\ 1}^{\ 2} \ , \ k_{\ 1}^{\ 2} \ , \ k_{\ 3}^{\ 2} \ ) \ \sim
\hspace*{0.2cm}
      \begin{array}{c} 
       g^{\ 2}
        \vspace*{0.2cm} \\
        \hline \vspace*{-0.2cm} \\
        12 \ \pi^{\ 2}
      \end{array}
\hspace*{0.2cm}
\ + \ \Pi \ \left \lbrack \ m_{\ Q} \ \overline{Q} \ Q \ \right \rbrack
\end{array}
\end{equation}

\noindent
In eq. (\ref{eq:4}) the dots represent all remaining covariant tensors 
( $O \ k^{\ 4}$ ) compatible with color gauge invariance.
$k_{\ 3} \ = \ k_{\ 2} \ - \ k_{\ 1}$ is the momentum transfer.
\vspace*{0.1cm} 

\noindent
$\Pi \ \left \lbrack \ m_{\ Q} \ \overline{Q} \ Q \ \right \rbrack$ 
denotes the contribution of the 'heavy' quark insertion alone.
The latter is readily calculated to the given order :

\begin{equation}
\begin{array}{l}
  \label{eq:5}
\Pi \ \left \lbrack \ m_{\ Q} \ \overline{Q} \ Q \ \right \rbrack \ =
\hspace*{0.2cm}
      \begin{array}{c} 
       g^{\ 2}
        \vspace*{0.2cm} \\
        \hline \vspace*{-0.2cm} \\
        8 \ \pi^{\ 2}
      \end{array} 
\hspace*{0.2cm}
       \Pi_{\ 1}^{\ (Q)}
        \vspace*{0.3cm} \\
\begin{array}{ll}
\Pi_{\ 1}^{\ (Q)} \ = & {\displaystyle{\int}} \ d \ \Omega_{\ 3} 
\ \left ( \ 4 \ \beta_{\ 1} \ \beta_{\ 2} \ - \ 1 \ \right ) \ \times 
        \vspace*{0.3cm} \\
& \times \ \left \lbrack
\hspace*{0.2cm}
      \begin{array}{c} 
       m^{\ 2}_{\ Q}
        \vspace*{0.2cm} \\
        \hline \vspace*{-0.2cm} \\
       X \ ( \ k_{\ 1}^{\ 2} \ , \ k_{\ 2}^{\ 2} \ , \ k_{\ 3}^{\ 2} \ )
      \end{array} 
\hspace*{0.2cm}
\ +
\hspace*{0.2cm}
      \begin{array}{c} 
       m^{\ 2}_{\ Q}
        \vspace*{0.2cm} \\
        \hline \vspace*{-0.2cm} \\
       X \ ( \ k_{\ 2}^{\ 2} \ , \ k_{\ 1}^{\ 2} \ , \ k_{\ 3}^{\ 2} \ )
      \end{array} 
\hspace*{0.2cm}
\ \right \rbrack
\end{array}
        \vspace*{0.5cm} \\
X \ ( \ k_{\ 1}^{\ 2} \ , \ k_{\ 2}^{\ 2} \ , \ k_{\ 3}^{\ 2} \ ) \ =
        \vspace*{0.3cm} \\
\hspace*{0.5cm}
\ = \ m^{\ 2}_{\ Q} \ - \ \left \lbrack
\ \beta_{\ 2} \ \beta_{\ 3} \ k_{\ 1}^{\ 2}
\ + \ \beta_{\ 1} \ \beta_{\ 3} \ k_{\ 2}^{\ 2}
\ + \ \beta_{\ 1} \ \beta_{\ 2} \ k_{\ 3}^{\ 2}
\ \right \rbrack
        \vspace*{0.3cm} \\
d \ \Omega_{\ 3} \ = \ \delta \ ( \ 1 \ - \ \sum_{\ k}^{\ 3} \ \beta_{\ k} \ ) 
\ \prod_{\ n}^{\ 3} \ \vartheta \ ( \ \beta_{\ n} \ ) \ d \ \beta_{\ n}
\end{array}
\end{equation}

\noindent
In the 'heavy' flavor limit the expression in eq. (\ref{eq:5}) becomes

\begin{equation}
\begin{array}{l}
  \label{eq:6}
\lim_{\ m_{\ Q} \ \rightarrow \ \infty}
\ \Pi \ \left \lbrack \ m_{\ Q} \ \overline{Q} \ Q \ \right \rbrack \ = \ -
\hspace*{0.2cm}
      \begin{array}{c} 
       g^{\ 2}
        \vspace*{0.2cm} \\
        \hline \vspace*{-0.2cm} \\
        12 \ \pi^{\ 2}
      \end{array} 
    \hspace*{0.3cm} \rightarrow \hspace*{0.1cm}
\begin{array}[t]{c}
\lim
        \vspace*{0.0cm} \\
_{\ m_{\ Q} \ \rightarrow \ \infty} 
\end{array}
\ \Pi \ = \ 0
        \vspace*{0.3cm} \\
2 \ {\displaystyle{\int}} \ d \ \Omega_{\ 3} 
\ \left ( \ 4 \ \beta_{\ 1} \ \beta_{\ 2} \ - \ 1 \ \right ) \ = \ - \ \frac{2}{3}
\end{array}
\end{equation}

\noindent
The quark part of the trace anomaly is revealed through the nontrivial
'heavy' mass insertion
$\left \lbrack \ m_{\ Q} \ \overline{Q} \ Q \ \right \rbrack$ in eq. (\ref{eq:6}).
\vspace*{0.1cm} 

\noindent
In a similar way we discuss the axial current anomaly, considering the two axial
currents

\begin{equation}
\begin{array}{l}
  \label{eq:7}
  a_{\ \mu}^{\ (+)} \ = \ \overline{q} \ \gamma_{\ \mu} \ \gamma_{\ 5} \ q
  \ - 
  \ n_{\ fl} \ \overline{Q} \ \gamma_{\ \mu} \ \gamma_{\ 5} \ Q
    \hspace*{0.3cm} ; \hspace*{0.3cm}
  a_{\ \mu} \ = \ \overline{q} \ \gamma_{\ \mu} \ \gamma_{\ 5} \ q
        \vspace*{0.3cm} \\
\gamma_{\ 5} \ = \ \gamma_{\ 5 \ R} \ = \ - \ i 
\ \gamma_{\ 0} \ \gamma_{\ 1} \ \gamma_{\ 2} \ \gamma_{\ 3} 
\end{array}
\end{equation}

\noindent
From the two currents defined in eq. (\ref{eq:7}) we form the divergences 

\begin{equation}
\begin{array}{l}
  \label{eq:8}
  D^{\ (+)} \ = \ \partial^{\ \mu}
\ a_{\ \mu}^{\ (+)} \ = 
\ 2 \ m_{\ q} \ \overline{q} \ i \ \gamma_{\ 5} \ q \ -
\ 2 \ n_{\ fl} 
\ m_{\ Q} \ \overline{Q} \ i \ \gamma_{\ 5} \ Q
        \vspace*{0.3cm} \\
  D \ = \ \partial^{\ \mu}
\ a_{\ \mu} \ = 
\ 2 \ m_{\ q} \ \overline{q} \ i \ \gamma_{\ 5} \ q \ +
\ 2 \ n_{\ fl} \ {\cal{B}} \ \widetilde{{\cal{B}}}
\hspace*{0.2cm}
      \begin{array}{c} 
        1
        \vspace*{0.2cm} \\
        \hline \vspace*{-0.2cm} \\
        8 \ \pi^{\ 2}
      \end{array}
\hspace*{0.2cm}
        \vspace*{0.3cm} \\
{\cal{B}} \ \widetilde{{\cal{B}}} \ =
\ {\cal{B}} \ \widetilde{{\cal{B}}} \ ( \ N_{\ c} \ ) \ =
     \ \left[
     \ \frac{1}{4} 
     \  F_{\mu \nu}^{a}
     \ \widetilde{F}^{a \, \mu \nu}
     \ \right]
    \hspace*{0.3cm} ; \hspace*{0.3cm}
\widetilde{F}^{a}_{\ \mu \nu} \ = \ \frac{1}{2} 
\ \varepsilon_{\ \mu \nu \sigma \tau}
     \  F^{a \ \sigma \tau}
\end{array}
\end{equation}

\noindent
and their difference

\begin{equation}
\begin{array}{l}
  \label{eq:9}
  \Delta \ D \ =
  \ D^{\ (+)} \ - \ D \ = \ - 2 \ n_{\ fl} \ \Delta^{\ (5)}
        \vspace*{0.3cm} \\
  \Delta^{\ (5)} \ = 
\ {\cal{B}} \ \widetilde{{\cal{B}}}
\hspace*{0.2cm}
      \begin{array}{c} 
        1
        \vspace*{0.2cm} \\
        \hline \vspace*{-0.2cm} \\
        8 \ \pi^{\ 2}
      \end{array}
\hspace*{0.2cm}
+ \ m_{\ Q} \ \overline{Q} \ i \ \gamma_{\ 5} \ Q
\end{array}
\end{equation}

\noindent
As in eq. (\ref{eq:4}) we evaluate the two gluon matrix element of
$\Delta^{\ (5)}$ in lowest order 

\begin{equation}
\begin{array}{l}
  \label{eq:10}
  \left \langle \ \varepsilon_{\ 2} \ , \ a_{\ 2} \ , \ k_{\ 2} \ \right |
\ \left ( \ \Delta^{\ (5)} \ \vartheta \ \right )^{\ \mu}_{\ \mu}
\ \left | \ \varepsilon_{\ 1} \ , \ a_{\ 1} \ , \  \ k_{\ 1} \ \right \rangle \ =
        \vspace*{0.3cm} \\
\delta_{\ a_{2} \ a_{1}} 
\ \varepsilon_{\ 2}^{\ * \ \sigma_{\ 2}} \ \varepsilon_{\ 1}^{\ \sigma_{\ 1}}
\ \varepsilon_{\ \sigma_{ 2} \ \sigma_{ 1} \ \mu_{ 2} \ \mu_{ 1}}
\ ( \ k_{ 2} \ )^{\ \sigma_{\ 2}} \ ( \ k_{ 1} \ )^{\ \sigma_{\ 1}}
\ \Pi^{\ (5)} \ + \ ...
        \vspace*{0.3cm} \\
\Pi^{ (5)} \ =
\ \Pi^{ (5)} \ ( \ k_{\ 1}^{\ 2} \ , \ k_{\ 1}^{\ 2} \ , \ k_{\ 3}^{\ 2} \ ) \ \sim
\, - \hspace*{0.1cm}
      \begin{array}{c} 
       g^{\ 2}
        \vspace*{0.2cm} \\
        \hline \vspace*{-0.2cm} \\
        8 \ \pi^{\ 2}
      \end{array}
\hspace*{0.1cm}
\ + \ \Pi^{ (5)} \ \left \lbrack \ m_{\ Q} \ \overline{Q} \ i \ \gamma_{\ 5} 
\ Q \ \right \rbrack
\end{array}
\end{equation}

\noindent
The notation is analogous to the one in eq. (\ref{eq:4}).
\vspace*{0.1cm} 

\noindent
The quantity $\Pi^{\ (5)} \ \left \lbrack \ m_{\ Q} \ \overline{Q} \ i \ \gamma_{\ 5} 
\ Q \ \right \rbrack$ now relates to the pseudoscalar mass insertion
corresponding to $m_{\ Q} \ \overline{Q} \ i \ \gamma_{\ 5} \ Q$. The 
pseudoscalar mass insertion is readily evaluated \cite{Adler} , in line
with the scalar mass insertion in eq. (\ref{eq:5})

\begin{equation}
\begin{array}{l}
  \label{eq:11}
\Pi^{\ (5)} \ \left \lbrack \ m_{\ Q} \ \overline{Q} \ i \ \gamma_{\ 5} 
\ Q \ \right \rbrack \ =
\hspace*{0.2cm}
      \begin{array}{c} 
       g^{\ 2}
        \vspace*{0.2cm} \\
        \hline \vspace*{-0.2cm} \\
        8 \ \pi^{\ 2}
      \end{array} 
\hspace*{0.2cm}
       \Pi^{\ (5) \ (Q)}_{\ 1}
        \vspace*{0.3cm} \\
\begin{array}{ll}
\Pi^{\ (5) \ (Q)}_{\ 1} &  = \ {\displaystyle{\int}} \ d \ \Omega_{\ 3} 
\ \times 
        \vspace*{0.3cm} \\
& \times \ \left \lbrack
\hspace*{0.2cm}
      \begin{array}{c} 
       m^{\ 2}_{\ Q}
        \vspace*{0.2cm} \\
        \hline \vspace*{-0.2cm} \\
       X \ ( \ k_{\ 1}^{\ 2} \ , \ k_{\ 2}^{\ 2} \ , \ k_{\ 3}^{\ 2} \ )
      \end{array} 
\hspace*{0.2cm}
\ +
\hspace*{0.2cm}
      \begin{array}{c} 
       m^{\ 2}_{\ Q}
        \vspace*{0.2cm} \\
        \hline \vspace*{-0.2cm} \\
       X \ ( \ k_{\ 2}^{\ 2} \ , \ k_{\ 1}^{\ 2} \ , \ k_{\ 3}^{\ 2} \ )
      \end{array} 
\hspace*{0.2cm}
\ \right \rbrack
\end{array}
\end{array}
\end{equation}

\noindent
The quantities $d \ \Omega_{\ 3}$ and X in eq. (\ref{eq:11}) 
are defined in eq. (\ref{eq:5}). Analogously to the scalar mass insertion 
in eq. (\ref{eq:6}) we find in the 'heavy' flavor limit

\begin{equation}
\begin{array}{l}
  \label{eq:12}
\lim_{\ m_{\ Q} \ \rightarrow \ \infty}
\ \Pi^{\ (5)} \ \left \lbrack \ m_{\ Q} \ \overline{Q} \ Q \ \right \rbrack \ = 
\hspace*{0.2cm}
      \begin{array}{c} 
       g^{\ 2}
        \vspace*{0.2cm} \\
        \hline \vspace*{-0.2cm} \\
        8 \ \pi^{\ 2}
      \end{array} 
    \hspace*{0.2cm} \rightarrow \hspace*{0.0cm}
\begin{array}[t]{c}
\lim
        \vspace*{0.0cm} \\
_{\ m_{\ Q} \ \rightarrow \ \infty} 
\end{array}
\ \Pi^{\ (5)} \ = \ 0
        \vspace*{0.3cm} \\
2 \ {\displaystyle{\int}} \ d \ \Omega_{\ 3} \ = \ 1
\end{array}
\end{equation}

\noindent
This concludes the illustration how single local 'heavy' quark operators
generate the quark induced parts of the central anomalies. This evidently
exhausts the axial current anomaly. The trace anomaly associated
with gauge boson self interaction - dominant for large $N_{\ c}$ -
can only be represented by 'extended heavy' flavors, fermions and (pseudo)scalars,
within the $N \ = \ 4$ susy theory. The corresponding 'heavy' mass terms of course
violate $N \ = \ 4$ supersymmetry. 
\vspace*{0.1cm} 

\noindent
{\bf 2. Mass square of $\eta^{\ \prime}$, double local operator insertions}
\vspace*{0.1cm} 

\noindent
To study the properties of $\eta^{\ \prime}$, we use the axial current
$a_{\ \mu}^{\ (+)}$ defined in eq. (\ref{eq:7}) and its divergence
$D^{\ (+)} \ = \ \partial^{\ \mu} a_{\ \mu}^{\ (+)}$ given in eq. (\ref{eq:8}).
In particular we consider the correlation function of {\em two} divergencies

\begin{equation}
\begin{array}{l}
  \label{eq:13}
  \chi^{\ (+)} \ ( \ q^{\ 2} \ ) \ =
\ i  {\displaystyle{\int}}  d^{\ 4} \ x \ e^{\ i \ q x}
\ \left \langle \ \Omega \ \right |
\ T \ \left \lbrack \ D^{\ (+)} \ ( \ x \ ) \ D^{\ (+)} \ ( \ 0 \ )
\ \right \rbrack
\ \left | \ \Omega \ \right \rangle
\end{array}
\end{equation}

\noindent
In the following we set all light quark masses to zero, when relevant.
The $SUn_{\ fl}$ singlet pseudoscalar meson contributes, eventually with
a finite width to $\chi^{\ (+)}$ which is of the general form

\begin{equation}
\begin{array}{l}
  \label{eq:14}
  \chi^{\ (+)} \ ( \ q^{\ 2} \ ) \ =
\ 2 \ n_{\ fl}
\hspace*{0.2cm}
      \begin{array}{c} 
        ( \ f^{\ (+)} \ )_{\ \eta^{\ \prime}}^{\ 2} \ ( \ q^{\ 2} \ )
        \ m_{\ \eta^{\ \prime}}^{\ 4} \ ( \ q^{\ 2} \ )
        \vspace*{0.2cm} \\
        \hline \vspace*{-0.2cm} \\
        m_{\ \eta^{\ \prime}}^{\ 2} \ ( \ q^{\ 2} \ )
        \ - q^{\ 2} 
      \end{array}
\hspace*{0.2cm}
\end{array}
\end{equation}

\noindent
In eq. (\ref{eq:14}) the quantities $m_{\ \eta^{\ \prime}}^{\ 2} \ ( \ q^{\ 2} \ )$
and $f^{\ (+)} \ )_{\ \eta^{\ \prime}}^{\ 2} \ ( \ q^{\ 2} \ )$ are extended 
to arbitrary values of $q^{\ 2}$, as off shell mass and decay constant (square)
of $\eta^{\ \prime}$. In the form given above, eq. (\ref{eq:14})
is exact for all values of $q^{\ 2}$. The Ward identity for the
axial current $a_{\ \mu}^{\ (+)}$ in eq. (\ref{eq:8}) implies

\begin{equation}
\begin{array}{l}
  \label{eq:15}
  \chi^{\ (+)} \ ( \ q^{\ 2} \ ) \ =
\vspace*{0.3cm} \\
 \left \lbrace \begin{array}{r}
q^{\ \mu} 
 {\displaystyle{\int}}  d^{\ 4} \ x \ e^{\ i \ q x}
 \left \langle \ \Omega \ \right |
\ T \ \left \lbrack \ a^{\ (+)}_{\ \mu}
\ ( \ x \ ) \ D^{\ (+)} \ ( \ 0 \ )
\ \right \rbrack
\ \left | \ \Omega \ \right \rangle
\vspace*{0.3cm} \\
- \ i \ 
\ {\displaystyle{\int}}  d^{\ 4} \ x \ e^{\ i \ q x} \ \delta \ ( \ t \ )
\ \left \langle \ \Omega \ \right |
\ \left \lbrack \ a^{\ (+)}_{\ 0} \ ( \ x \ ) \ , \ D^{\ (+)} \ ( \ 0 \ )
\ \right \rbrack
\ \left | \ \Omega \ \right \rangle
\end{array}
 \right \rbrace
\end{array}
\end{equation}

\noindent
The equal time commutator in eq. (\ref{eq:15}) involves the 'heavy' flavor
only. It gives rise to the induced contact term, which I have discussed earlier 
\cite{PMeta}, yielding a remnant contribution, persisting when we
evaluate $\chi^{\ (+)}$ at zero momentum :

\begin{equation}
\begin{array}{l}
  \label{eq:16}
  \chi^{\ (+)} \ ( \ 0 \ ) \ =
\ 4 \ n_{\ fl}^{\ 2}
\ \left \langle \ \Omega \ \right |
\ \left ( \ - \ m_{\ Q} \ \overline{Q} \ Q \ \right )
\ \left | \ \Omega \ \right \rangle
\end{array}
\end{equation}

\newpage 

\noindent
In the 'heavy' flavor limit it follows, combining
eqs. (\ref{eq:3}) , (\ref{eq:14}) and (\ref{eq:16})

\begin{equation}
\begin{array}{l}
  \label{eq:17}
f_{\ \eta^{\ \prime}}^{\ 2} \ ( \ 0 \ )
\ m_{\ \eta^{\ \prime}}^{\ 2} \ ( \ 0 \ )
\ =
\hspace*{0.2cm}
      \begin{array}{c} 
        n_{\ fl}
        \vspace*{0.2cm} \\
        \hline \vspace*{-0.2cm} \\
        3
      \end{array}
\hspace*{0.2cm}
      \begin{array}{c} 
        1
        \vspace*{0.2cm} \\
        \hline \vspace*{-0.2cm} \\
        2 \ \pi^{\ 2}
      \end{array}
\hspace*{0.2cm}
 \left \langle \ \Omega \ \right |
\ {\cal{B}}^{\ 2}
\ \left | \ \Omega \ \right \rangle
\vspace*{0.3cm} \\
f_{\ \eta^{\ \prime}}^{\ 2} \ ( \ q^{\ 2} \ ) \ =
\ \begin{array}[t]{c}
\lim
        \vspace*{0.0cm} \\
_{\ m_{\ Q} \ \rightarrow \ \infty} 
\end{array}
        \ ( \ f^{\ (+)} \ )_{\ \eta^{\ \prime}}^{\ 2} \ ( \ q^{\ 2} \ )
\end{array}
\end{equation}

\noindent
The relation in eq. (\ref{eq:17}) reveales the 'heavy' quark contact term
as an anomalous contribution involving the insertion of two local operators.
This affects the topological susceptibility through nontrivial boundary conditions

\begin{equation}
\begin{array}{l}
  \label{eq:18}
 \begin{array}[t]{c}
\lim
        \vspace*{0.0cm} \\
_{\ m_{\ Q} \ \rightarrow \ \infty} 
\end{array}
\ \chi^{\ (+)} \ ( \ 0 \ ) \ =
        \vspace*{0.3cm} \\
\hspace*{0.2cm}
 4 \ n_{\ fl}^{\ 2} \ i
\ {\displaystyle{\int}}  d^{\ 4} \ x
\ \left \langle \ \Omega \ \right |
\ T \ \left \lbrack \ ch_{\ 2} \ ( \ x \ )
\ ch_{\ 2} \ ( \ 0 \ )
\ \right \rbrack
\ \left | \ \Omega \ \right \rangle
\vspace*{0.3cm} \\
ch_{\ 2} \ =
\ {\cal{B}} \ \widetilde{{\cal{B}}}
\hspace*{0.2cm}
      \begin{array}{c} 
        1
        \vspace*{0.2cm} \\
        \hline \vspace*{-0.2cm} \\
        8 \ \pi^{\ 2}
      \end{array}
\end{array}
\end{equation}

\noindent
Eq. (\ref{eq:17}) falsifies the pertubatively motivated hypothesis  
\cite{err} \footnote{We refrain from citing here any of the numerous more recent
papers, expanding on the incorrect large $N_{\ c}$ limit.}

\begin{equation}
\begin{array}{l}
  \label{eq:19}
m_{\ \eta^{\ \prime}}^{\ 2} \ \propto \ 1 \ / \ N_{\ c} 
\end{array}
\end{equation}

\noindent
in the large $N_{\ c}$ limit, since both 
the gauge boson condensate $ \left \langle \ \Omega \ \right | \ {\cal{B}}^{\ 2}
\ \left | \ \Omega \ \right \rangle$ and the square of the 
$\eta^{\ \prime}$ decay constant $f_{\ \eta^{\ \prime}}^{\ 2} \ ( \ 0 \ )$
are proportional to $N_{\ c}$ in that limit.
\vspace*{0.1cm} 

\noindent
Thus the extrapolation in $q^{\ 2}$ from
$m_{\ \eta^{\ \prime}}^{\ 2}$ to zero does not allow a straightforward
identification of 
$m_{\ \eta^{\ \prime}} \ ( \ 0 \ )$ with the physical mass of
$\eta^{\ \prime}$ nor of
$f_{\ \eta^{\ \prime}} \ ( \ 0 \ )$ with its decay constant.
\vspace*{0.1cm} 

\noindent
We rescale the remnant relations combining eqs. (\ref{eq:16}) and (\ref{eq:18})
in the 'heavy' flavor limit to the topological susceptibility per unit winding
number $\chi_{\ ch}$

\begin{equation}
\begin{array}{l}
  \label{eq:20}
 \begin{array}[t]{c}
\lim
        \vspace*{0.0cm} \\
_{\ m_{\ Q} \ \rightarrow \ \infty} 
\end{array}
\ \chi^{\ (+)} \ ( \ 0 \ ) \ = \ 4 \ n_{\ fl}^{\ 2} \ \chi_{\ ch}
    \hspace*{0.3cm} ; \hspace*{0.3cm}
    \Theta_{\ F} \ = 
\ {\cal{B}}^{\ 2}
\hspace*{0.2cm}
      \begin{array}{c} 
        1
        \vspace*{0.2cm} \\
        \hline \vspace*{-0.2cm} \\
        12 \ \pi^{\ 2}
      \end{array}
        \vspace*{0.3cm} \\
\chi_{\ ch}  =
 i \ {\displaystyle{\int}}  d^{\ 4} \ x
\ \left \langle \ \Omega \ \right |
 T \ \left \lbrack \ ch_{\ 2} \ ( \ x \ )
\ ch_{\ 2} \ ( \ 0 \ )
\ \right \rbrack
\ \left | \ \Omega \ \right \rangle
 =
 \left \langle \ \Omega \ \right |
    \ \Theta_{\ F}
\ \left | \ \Omega \ \right \rangle
\end{array}
\end{equation}

\noindent
$\chi_{\ ch}$ defined in eq. (\ref{eq:20}) is to be evaluated retaining {\em all}
light flavors of gauge bosons {\em and} quarks.
\vspace*{0.1cm} 

\noindent
The Euclidean space version of eq. (\ref{eq:20})
in a finite large fourdimensional volume $V \ = \ V_{\ 4}$ becomes using
thermodynamic notation \cite{Crew} 
$\chi_{\ ch} \ \rightarrow ( \ \Delta \ \nu \ )^{\ 2} \ ( \ V \ ) \ / \ V$

\begin{equation}
\begin{array}{l}
  \label{eq:21}
  ( \ \Delta \ \nu \ )^{\ 2} \ ( \ V \ ) \ = \ N_{\ F} \ ( \ V \ ) 
  \ \rightarrow \ \varrho_{\ F} \ V
        \vspace*{0.3cm} \\
  \varrho_{\ F} \ = 
\ \left \langle \ \Omega \ \right |
    \ \Theta_{\ F}
\ \left | \ \Omega \ \right \rangle \ \propto \ N_{\ c}
\end{array}
\end{equation}

\noindent
There is a definite difference in sign - not an error - in the definition of
the quantities $\chi_{\ ch}$ and $\Delta \ \nu^{\ 2}$
in eqs. (\ref{eq:20}) and (\ref{eq:21}) relative to ref. \cite{Crew} . Which
correlation function is positive is controlled here by the 'heavy' flavor
and the sign of the gauge boson condensate. The sign in question reveales
an interesting property of Euclidean relative time correlations
with respect to their physical time counterparts.
\vspace*{0.1cm} 

\noindent
We conclude by evaluating, despite the large extrapolation, 
the relation in eq. (\ref{eq:17})
using the value of the gauge boson condensate derived by Shifman, Vainshtain
and Zakharov from charmonium sum rules \cite{SVZ}, the pion
decay constant for $f_{\ \eta^{\ \prime}} \ ( \ 0 \ )$
and the physical mass of 
$\eta^{\ \prime}$ for $m_{\ \eta^{\ \prime}} \ ( \ 0 \ )$

\begin{equation}
\begin{array}{l}
  \label{eq:22}
\left \langle \ \Omega \ \right | \ {\cal{B}}^{\ 2}
\ \left | \ \Omega \ \right \rangle \ \sim \ 0.125 \ \mbox{GeV}^{\ 4}
\vspace*{0.3cm} \\
f_{\ \eta^{\ \prime}} \ ( \ 0 \ ) \ \sim \ f_{\ \pi} \ = 0.0932 \ \mbox{GeV}
\end{array}
\end{equation}
%


\noindent
We find

\begin{equation}
\begin{array}{l}
  \label{eq:23}
m_{\ \eta^{\ \prime}}^{\ 2} \ \sim 
\hspace*{0.2cm}
      \begin{array}{c} 
        n_{\ fl}
        \vspace*{0.2cm} \\
        \hline \vspace*{-0.2cm} \\
        3
      \end{array}
\hspace*{0.2cm}
0.73 \ \mbox{GeV}^{\ 2}
    \hspace*{0.3cm} ; \hspace*{0.3cm}
m_{\ \eta^{\ \prime}}^{\ 2} \ - \ \frac{2}{3} \ m_{\ K}^{\ 2} \ = \ 0.75 
\ \mbox{GeV}^{\ 2}
\end{array}
\end{equation}

\noindent
For comparison we reduced the physical value of 
$m_{\ \eta^{\ \prime}}^{\ 2}$ by $\frac{2}{3} \ m_{\ K}^{\ 2}$ to offset the 
nonvanishing mass of the strange quark, ignoring $\eta$ $\eta^{\ \prime}$ 
mixing. Taking the latter into account improves  the numerical agreement 
beyond the theoretical error in the gauge boson condensate.


\vspace*{0.3cm} 

\noindent
\large{\bf Acknowledgments}
\vspace*{0.1cm} 
 
\noindent
I should like to thank C, Greub and M. Leibundgut for raising pertinent 
questions and for the ensuing discussions.











\end{document}